\newif\ifarxiv
\arxivtrue % or
\ifarxiv
	\documentclass[12pt]{article}
	\usepackage[total={6.5in, 9in}]{geometry}
	\usepackage{tgtermes}
	\usepackage[slantedGreek]{mathpazo}
\else
	\documentclass[conference]{IEEEtran}
\fi

\usepackage{graphicx}
\usepackage{float}
\usepackage{amsmath}
\usepackage{algorithmic}
\usepackage{array}
\usepackage{color}
\usepackage{url}

\graphicspath{{./fig/}}
\begin{document}
\ifarxiv
\author{
	Vadim Sokolov\footnote{Corresponding author, email: vsokolov@gmu.edu} and  Muhammad Imran\\ \textit{George Mason University}\\
	David W. Etherington\\\textit{Connected Signals, Inc.}\\
	Christian Schmid\\\textit{BMW North America, Inc.}\\
	Dominik Karbowski and Aymeric Rousseau\\\textit{Argonne National Laboratory}
}
\title{Effects of Predictive Real-Time Traffic Signal Information}%\thanksref{T1}}
\date{First Draft: December 2017\\
	This Draft: November 2018
}
\maketitle
\else
\title{~\\~\\Effects of Predictive Real-Time Traffic Signal Information}
\author{
\IEEEauthorblockN{Vadim Sokolov, Muhammad Imran}
\IEEEauthorblockA{%Systems Engineering and\\Operations Research\\
George Mason University\\
Fairfax, VA\\
Email: \{vsokolov,mimran@gmu.edu\}}
\and
\IEEEauthorblockN{David W. Etherington}
\IEEEauthorblockA{Connected Signals, Inc.\\
	Eugene, OR \\
	Email: ether@connectedsignals.com}
\and
%\IEEEauthorblockN{Christian Schmid}
%\IEEEauthorblockA{BMW North America, Inc.\\
%Mountain View, CA \\
%	Email: Christian.CA.Schmid@bmw.de}
%\and
%\IEEEauthorblockN{Dominik Karbowski}
%\IEEEauthorblockA{Argonne National Laboratory\\
%	Lemont, IL\\
%	Email: dkarbowski@anl.gov}
\and
\IEEEauthorblockN{Dominik Karbowski, Aymeric Rousseau}
\IEEEauthorblockA{Argonne National Laboratory\\
Lemont, IL\\
Email: \{dkarbowski, arousseau\}@anl.gov}
}
\maketitle
\fi	

\begin{abstract}
This paper analyzes the impact of providing car drivers with predictive information on traffic signal timing in real-time, including time-to-green and green-wave speed recommendations. Over a period of six months, the behavior of these 121 drivers in everyday urban driving was analyzed with and without access to live traffic signal information. In a first period, drivers had the information providing service disabled in order to establish a baseline behavior; after that initial phase, the service was activated. In both cases, data from smartphone and vehicle sensors was collected, including speed, acceleration, fuel rate, acceleration and brake pedal positions. We estimated the changes in the driving behavior which result from drivers' receiving the traffic signal timing information by carefully comparing distributions of acceleration/deceleration patterns through statistical analysis. Our analysis demonstrates that there is a positive effect of providing traffic signal information timing to the drivers.

%For negative acceleration values, the Kolmogorov-Smirnov $D$ statistic was equal to 0.16112, with the  $p$-value $<10^{-16}$. For negative acceleration values, the Kolmogorov-Smirnov $D$ statistic was equal to 0.029706, with the  $p$-value $=1.8\times 10^{-6}$. 
\end{abstract}

% \fi
\ifarxiv
\else
\IEEEpeerreviewmaketitle
\fi

\section{Introduction}
Vehicle-to-Infrastructure (V2I) systems have been the subject of intense interest in recent years, offering the promise of significant reductions in fuel consumption and greenhouse gas and other emissions, as well as safety improvements. While there have been many testbed studies that support these promises, as well as anecdotal evidence from limited deployments, the effectiveness of various V2I mechanisms in real-world has not been fully demonstrated. Many factors that could impact, and possibly negate the anticipated benefits, such as driver compliance, distraction, and the impact of other drivers' behavior are difficult to estimate before wide-scale trials. In order to increase transportation authorities' willingness to deploy V2I systems, it is important that real-world studies be conducted to gather data that can serve to evaluate these factors. Such studies would, ideally, compare a statistically significant number of individual drivers' performance with and without V2I assistance, under a variety of driving conditions, for long enough to explore ``novelty" effects (e.g., whether drivers stop paying attention once they become habituated to the technology). 

One example of V2I technology consists in providing real-time information about upcoming traffic signals, which could bring 8-15\% energy savings \cite{cite5,cite6}. However, most studies have focused on modeling and simulation \cite{cite10,cite11} or ``professional driver on closed course" studies \cite{cite6}, which may not capture real-world complex factors. This technology could also reduce the number of accidents at signalized intersections. However, only real-world implementation would demonstrate whether that would be the case, and would also uncover any potential unintended consequences - technology supposed to improve safety, such as red-light cameras, sometimes produce counterintuitive results \cite{cite1}.

To help achieve the anticipated benefits, DSRC (Digital Short-Range Communications) is a possible communication technology that would allow the infrastructure to communicate with approaching vehicles, using specialized roadside and in-vehicle equipment. While DSRC offers many benefits, including nearly instantaneous relay of information, this approach requires a significant investment in new infrastructure.

Connected Signals has developed and demonstrated a complementary approach to relaying signal information to vehicles that exploits existing connections between traffic signals and municipal traffic management systems (TMSs), and existing connections between TMSs and the Internet, to access signal data. Cellular technology is used to communicate with vehicles. This approach avoids the need to deploy special-purpose hardware at each intersection and in each vehicle. A number of pilot deployments have been completed in cities in the US, New Zealand, and Australia. Given that a large fraction of urban traffic signals are connected to TMSs, and that vehicles increasingly have built-in cellular connectivity, this approach offers the prospect of being able to connect many signals to many vehicles almost immediately at very low cost. Signal information can be accessed through Connected Signals' EnLighten$^\mathrm{\textregistered}$~smartphone app, as well as directly through integrated systems that have been developed with a number of major vehicle OEMs, including BMW.

\section{Study Design}
\begin{figure}[H]
	\includegraphics[width=1\linewidth]{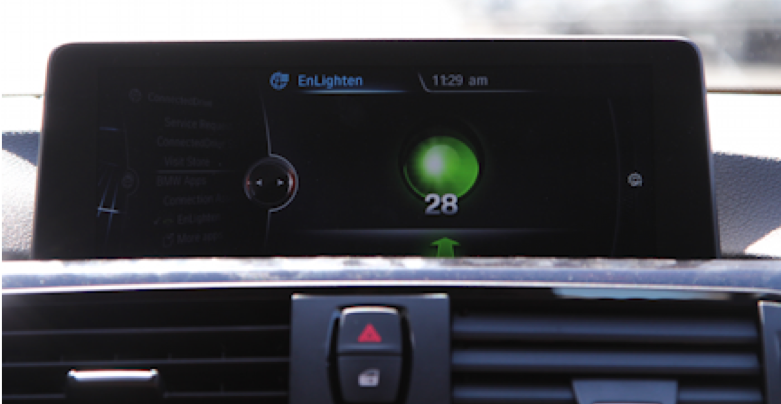}
	\caption{BMW EnLighten display: next light will be green, based on current speed}
	\label{fig:enlighten}
\end{figure}

In this study, Connected Signals (CS) data was broadcast to the drivers in one of two forms: through the vehicle infotainment unit for the drivers of BMWs equipped with the functionality, or through CS' EnLighten smartphone app. Special versions of each were produced to accommodate the study's needs. The initial fielded version supports signal count-downs and ``green-light-assist" information that tells drivers whether they would make an upcoming signal at their current speed. A sample screenshot is shown in Figure~\ref{fig:enlighten}. The display indicates that the light is currently green and will remain so for 28 seconds. The green arrow shows that, at the current speed, the car will arrive at the intersection during the green phase. The application is available in selected cities, including San Jose, CA, where the field study was conducted.

The study involved recruiting roughly 400 drivers. With the drivers' consent, data on vehicle position and speed, acceleration, braking, and (where possible) fuel consumption was collected. This data was transmitted to servers in the cloud, where it was merged with contemporaneous signal-state information for later, offline, analysis.

Drivers who spend significant portions of their driving time both in and out of the covered areas were recruited. This allows their behavior to be compared longitudinally, making it possible to detect habituation effects and eliminate biases that might occur in simple sequential ``without data/with data" trials.  Data collection was run for approximately six months, to ensure that a meaningful amount of data was collected for each driver, and that each driver experienced a variety of driving conditions.

When in covered areas, drivers were provided with predictive signal information telling them, when possible, whether they would make or miss the next signal at their current speed, a recommended speed to make the next signal, and countdowns for red signal durations when they are stopped. For safety reasons, speed recommendations were limited by the current speed limit, and red-light countdowns stopped at 5 seconds before the signal changes to force drivers to rely on the physical traffic signal. At that time, a chime also sounded to alert drivers to return their focus to driving in case they may have become distracted while waiting for the signal to change.

The experimental design for the study is intended to maximize our ability to determine the effects of signal data availability, given the constraints of what can readily be obtained from a collection of privately owned vehicles and a self-nominated group of participants.

A number of steps were taken to minimize-as much as possible-the effects of such factors as driver and vehicle variability, habituation, and differing driving conditions in and out of signal coverage. First of all, during an initial period, drivers were not  provided with signal information over a sufficient number of trips to establish a baseline. During that time, information was collected on drives and correlated with real-time signal state information. This allows determination of how drivers respond to the signal state information they get in the normal way (looking at the lights) without additional predictive or guidance information.

Secondly, throughout the study, data was collected from trips both inside and outside the signal-coverage area. Since the locations (but not the states) of signals outside of coverage are known, this helps distinguish between changes in drivers' behavior that result from access to signal data and changes that result from other factors such as weather or traffic conditions. While this is not a perfect comparison, it should provide reasonable indicators of the significance of the observed results. 

Finally, each driver was  assigned a unique ID that was used to associate all their drives. This allows changes in driver behavior to be analyzed longitudinally over the course of the study, including between control and signal-informed driving conditions. The unique IDs were created so they cannot be inverted to identify particular drivers to ensure the privacy of drivers in the study, and all data was  anonymized using these IDs as it was received.

Although the characteristics of the study's drivers and their route selections cannot be controlled for, the ability to compare individual drivers longitudinally, both with and without signal data, over an extended period should minimize the influence of such factors. 

For those vehicles with integrated signal information capabilities, the necessary information was captured directly from the participating vehicles. For vehicles without direct integration, a special-purpose version of EnLighten was  developed to acquire the necessary data from the smartphones' sensors. For each trip, time-series data was be collected on:

\begin{enumerate}
	\item Vehicle position, heading, and speed
	\item Number and duration of stops
	\item Acceleration and deceleration profiles
	\item Energy consumption (if available)
	\item Availability, timing, and content of provided signal information
	\item Actual signal state (if known).
\end{enumerate}

This information was associated with the vehicle type and driver ID. All data was sent to the cloud both to facilitate provision of signal-state predictions to the vehicle and for recording for subsequent analysis. 

Since baseline and longitudinal data without signal provisioning was collected in addition to data with signal information, it should be possible to reliably estimate the effects on energy consumption and safety and estimate the impact of signal time information.

\section{San Jose Data}
\begin{figure}[h]
	\begin{tabular}{cc}
		\includegraphics[width=0.45\linewidth]{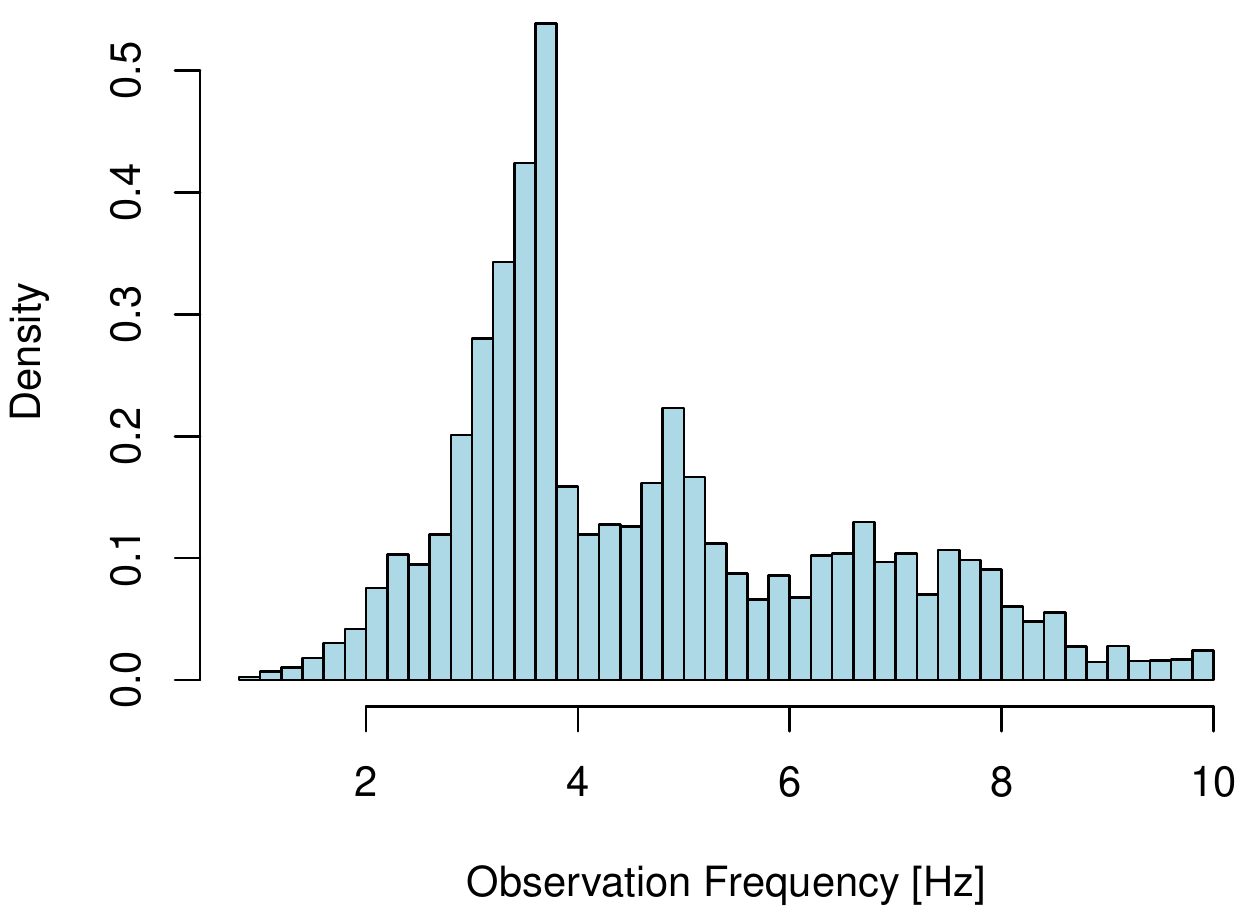} & \includegraphics[width=0.45\linewidth]{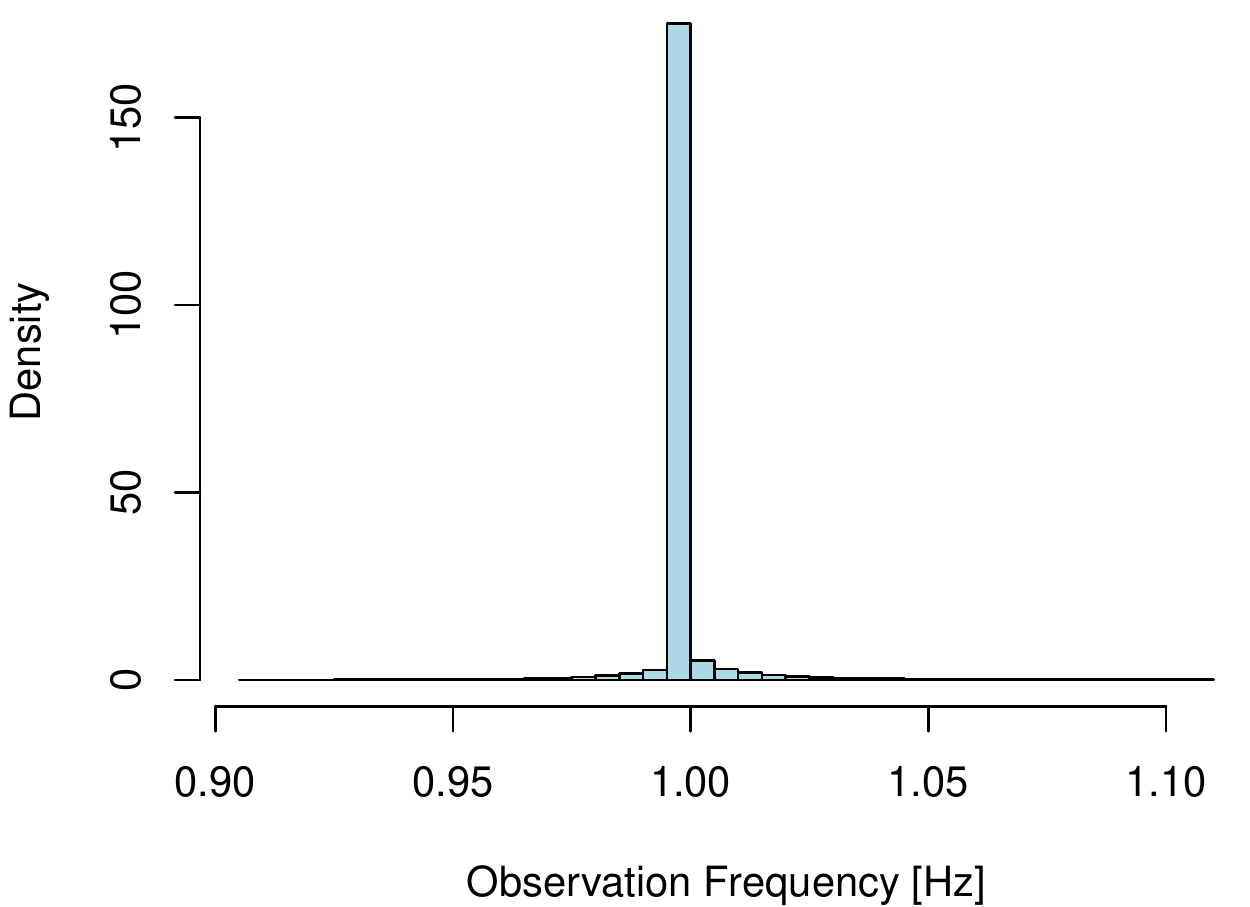}
	\end{tabular}
	\caption{Histogram of observations' frequency from (a) CAN bus and (b) phone sensors}
	\vspace{-10pt}
	\label{fig:freq}
\end{figure}

We analyze data from two sources: smartphone and CAN bus. The CAN bus data are collected via OBDII interface. The data was observed during the period from 2016-09-13 to 2017-02-09. Table~\ref{tab:nrec} provides number of observations (in thousands) and number of active and inactive trips from CAN and phone. An active trip is when the feedback system was on. Phone observations were collected from GPS and accelerometer sensors. CAN signals are vehicle dependent. Different manufacturers broadcast different sets of signals via the OBDII interface.  
\begin{table}[H] \centering 
	\begin{tabular}{lllll} 
		Dataset & Obs (A)$/ 10^3$  & Obs (I)$/ 10^3$ & Trips (A) & Trips (I)\\ 
		\hline \\[-1.8ex] 
		CAN Speed & 6461 & 9941 & 2535 & 4406 \\ 
		Phone Speed & 3172 & 3753 & 4661 & 6803 \\ 
		HMM & 2408 & 2918 & 2961 & 4359 \\ 
		Phone Acceleration & 2852 & 3861 & 4733 & 8033 \\ 
		CAN Acc. Pedal D & 2171 & 3325 & 2306 & 3911 \\ 
		CAN RPM & 1088 & 1726 & 2482 & 4315 \\ 
		CAN Throttle & 1088 & 1727 & 2450 & 4249 \\ 
		CAN Acc. Pedal E & 1085 & 1662 & 2307 & 3904 \\ 
		CAN Throttle R & 1068 & 1636 & 2299 & 3844 \\ 
		CAN Throttle B & 1051 & 1591 & 2268 & 3754 \\ 
		CAN Fuel Rate& 19 & 62 & 79 & 165 \\ 
		BMW RPM & 998 & 488 & 214 & 57 \\ 
		BMW Fuel & 916 & 460 & 214 & 57 \\ 
		BMW Acceleration & 9 & 171 & 41 & 13 \\ 
		BMW Location & 73 & 44 & 165 & 56 \\ 
		BMW Start/Stop & 13 & 2 & 214 & 57 \\ 
		BMW Brake & 9 & 0 & 41 & 3 \\ 
		Feedback Time & 414 & 0 & 2570 & 0 \\ 
		Feedback Green & 152 & 0 & 3303 & 0 \\ 
	\end{tabular} 
	\caption{Number of active (A) and inactive (I) observations and trips, Number of observations are in thousands.} 
	\label{tab:nrec} 
\end{table} 
Observations collected from CAN bus data is collected at irregular frequency, with most of the observations being collected at frequency of 3 Hz. On the other hand, phone data was observed at regular frequency of 1 Hz. Figure~\ref{fig:freq} shows the empirical distribution of the frequencies at which data was collected from both sources.

The irregularity of the CAN observations' frequency is most likely due to WiFi connection disruptions. The data from ODBII dongle was transmitted to the phone via WiFi connection and then the phone would send data to the back-end server.

\begin{figure}[H]
	\centering
	\includegraphics[width=0.9\linewidth]{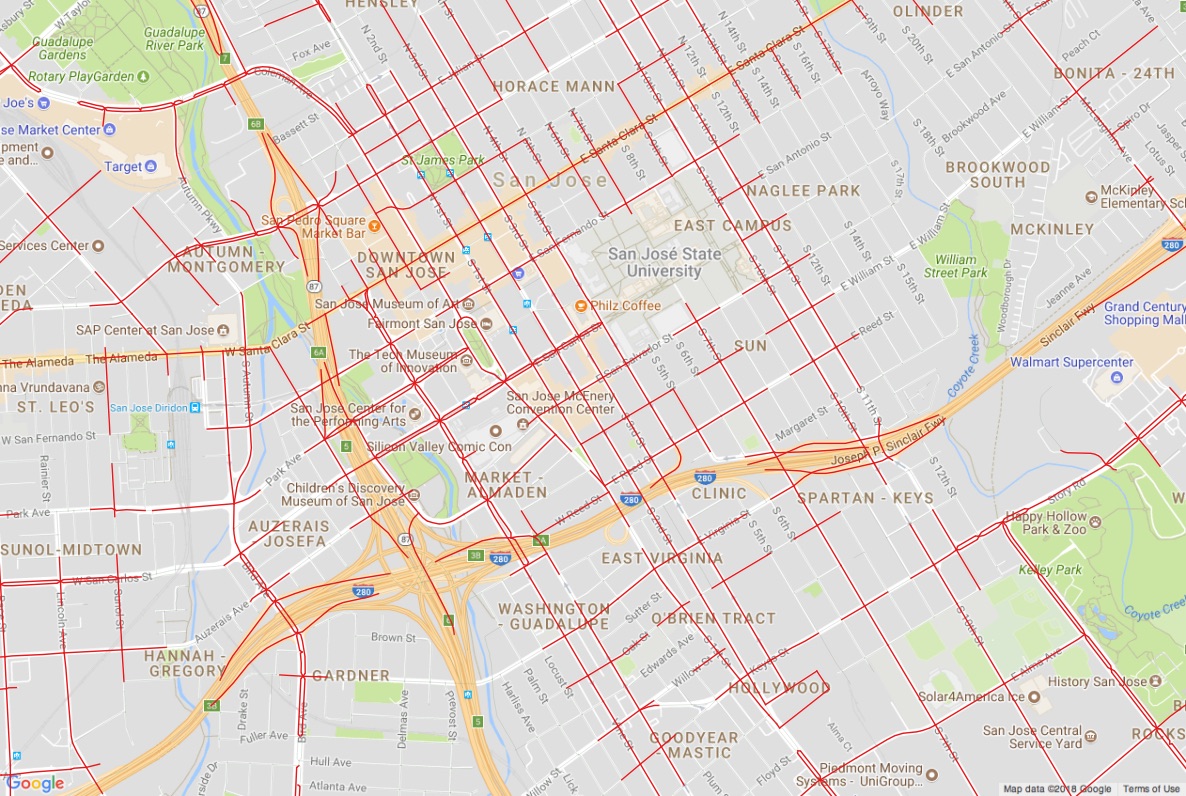}
	\caption{Active Road Segments. Road segments colored in {\color{red}red} are active}
	\label{fig:map-active}
\end{figure}

There are 13154 road segments from which the data was observed. Out of those, 3620 were road segments on which drivers would receive information about traffic light timing. Road segments on which traffic light information is available we will call \textbf{active segments}. As shown in Figure~\ref{fig:map-active}, most of the road segments in the central part of San Jose are active. 

\begin{figure}[h]
	\begin{tabular}{cc}
		\includegraphics[width=0.45\linewidth]{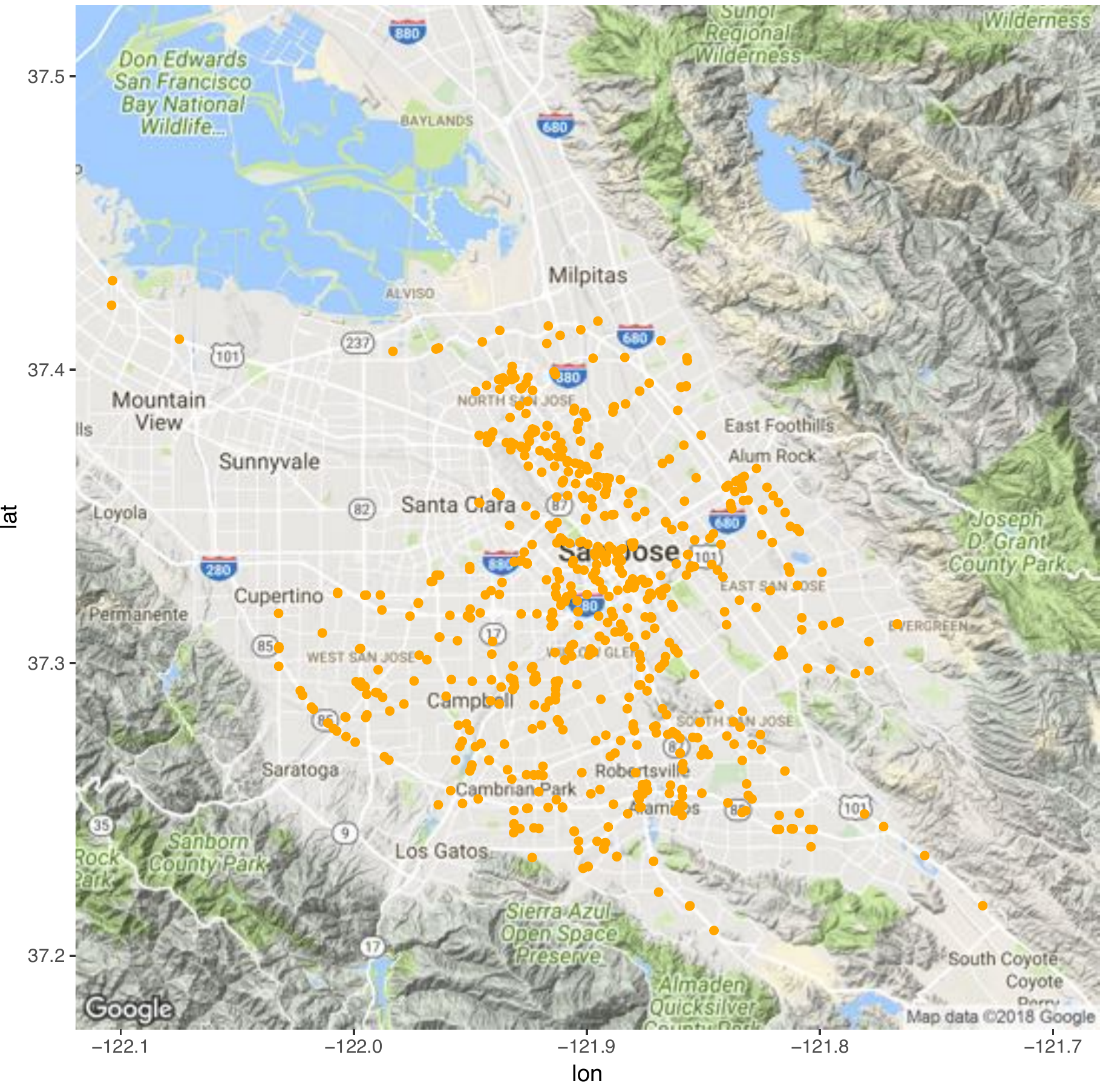} & \includegraphics[width=0.45\linewidth]{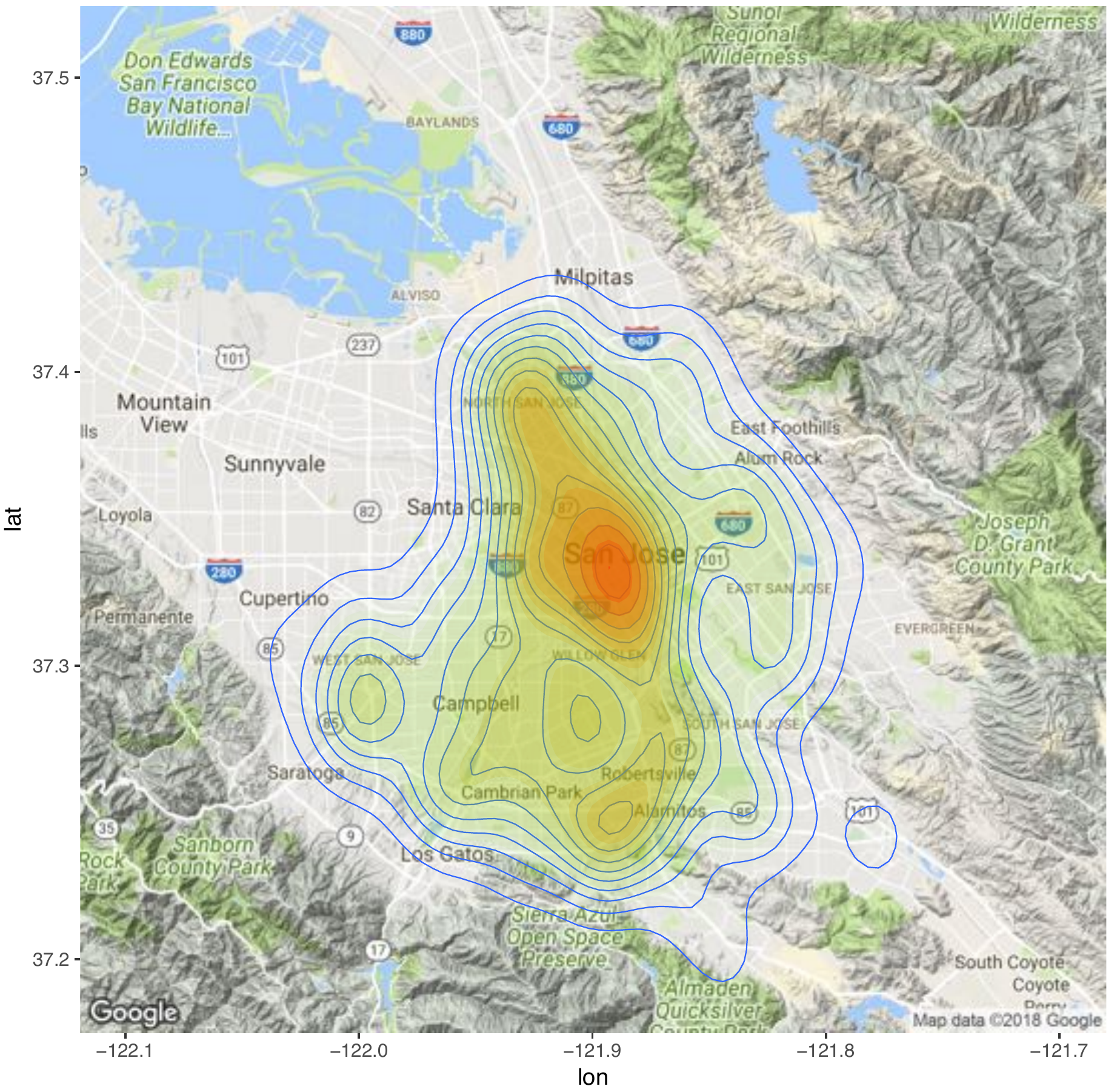}\\
		(a) Sampled Locations & (b) Heat map
	\end{tabular}
	\caption{Sample of locations where data was recorded}
	\vspace{-10pt}
	\label{fig:map-phone}
\end{figure}

Further, most of the data was collected in the central part of San Jose. Maps on Figure~\ref{fig:map-phone} show a 0.01\% sample of observations from the analyzed data set. 

\subsection{Data Processing}
\begin{figure}[h]
	\begin{tabular}{cc}
		\includegraphics[width=0.47\linewidth]{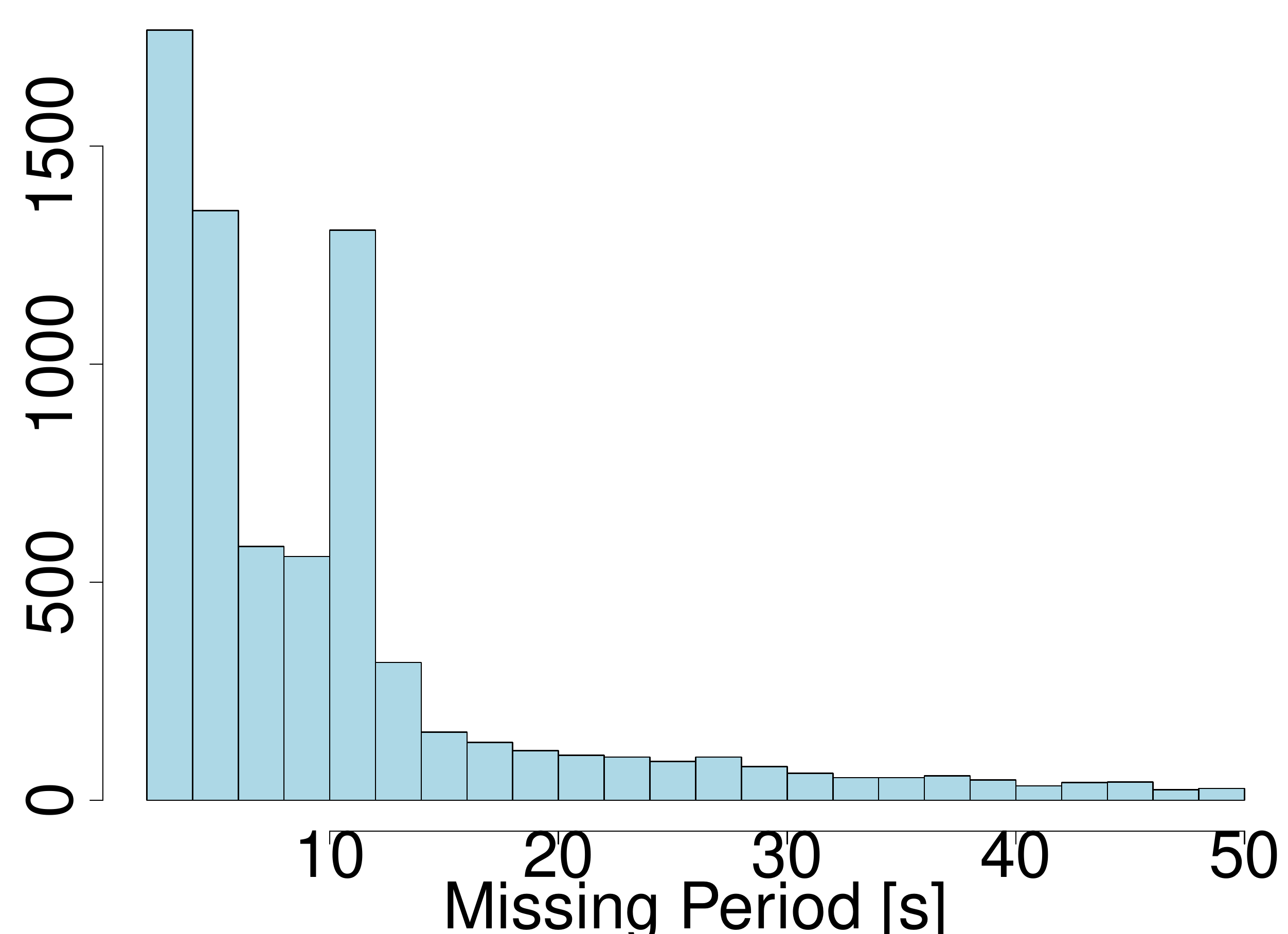} & \includegraphics[width=0.47\linewidth]{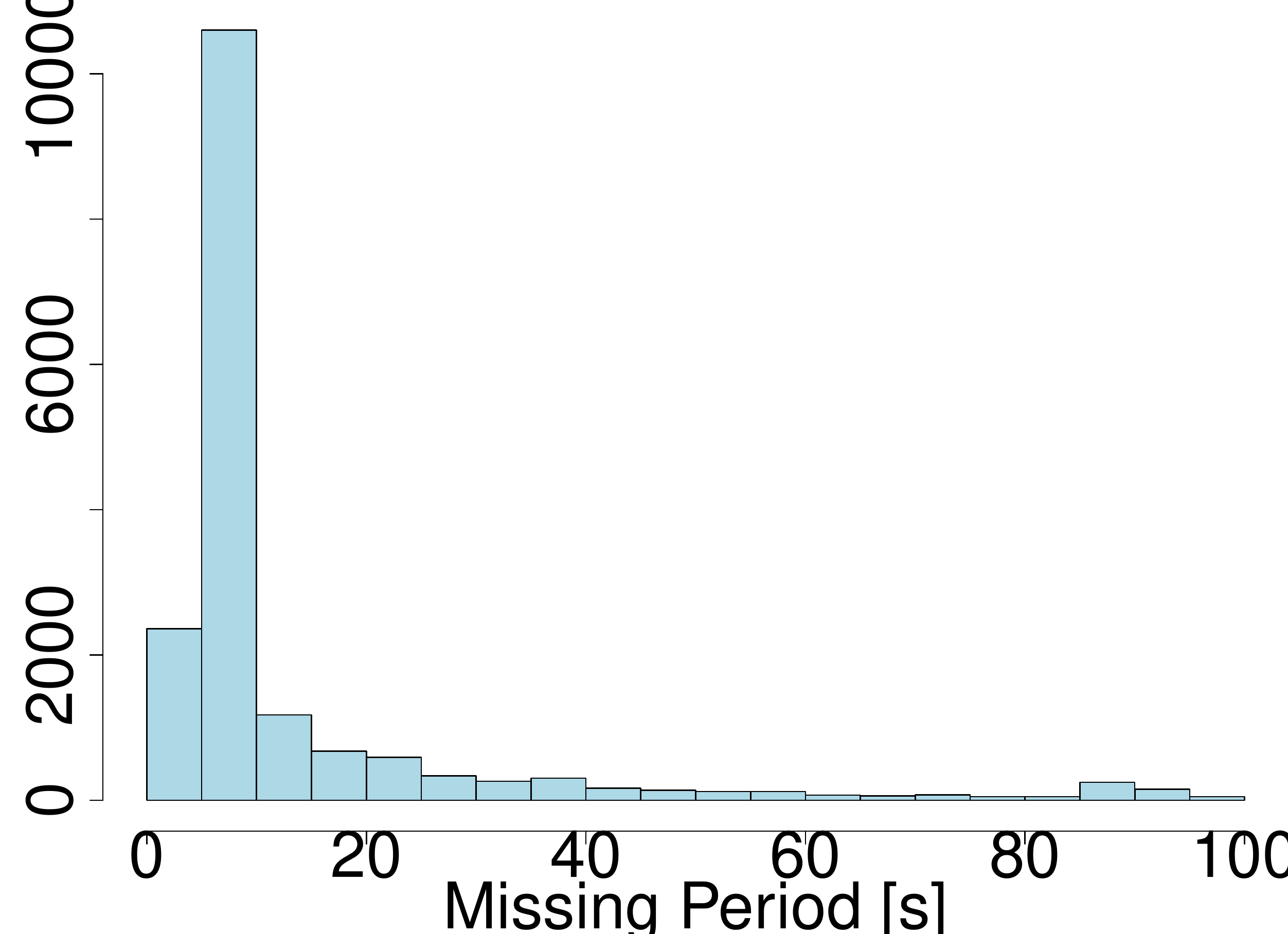}\\
		(a) CAN & (b) Phone
	\end{tabular}
	\caption{Histogram of duration of missing data from CAN and phone.}
	\vspace{-10pt}
	\label{fig:missing_data}
\end{figure}

Data from phone's GPS sensor was matched to road segments using Hidden Markov Model (HMM)~\cite{newson2009hidden,luo2015addressing}. Missing observations were considered to be at random. Plot~\ref{fig:missing_data} shows distribution of durations of missing observations. There is a heavy tail for durations of missing observations among CAN signals. This is likely due to the connectivity issues between OBDII dongle and the phone that collects the data. 

When we calculate derivatives of location to calculate the speed, we simply remove the observation with time jumps. We also observed that some of the trips had long sequence of zero speed observations. We removed those zero observations from analysis. 

To address the issue of noise we truncated observations with values that are beyond physical limits. Speed was truncated to be inside $[0,160]~mi/h$ miles per hour and acceleration was truncated to be inside $[-6,4]~m/s^2$. Further we used Kalman smoothing to remove sharp acceleration spikes and changes in speed that violate basic laws of vehicle dynamics.

\subsubsection{Kalman Smoothing}
\begin{figure}[H]
	\begin{tabular}{cc}
		\includegraphics[width=0.45\linewidth]{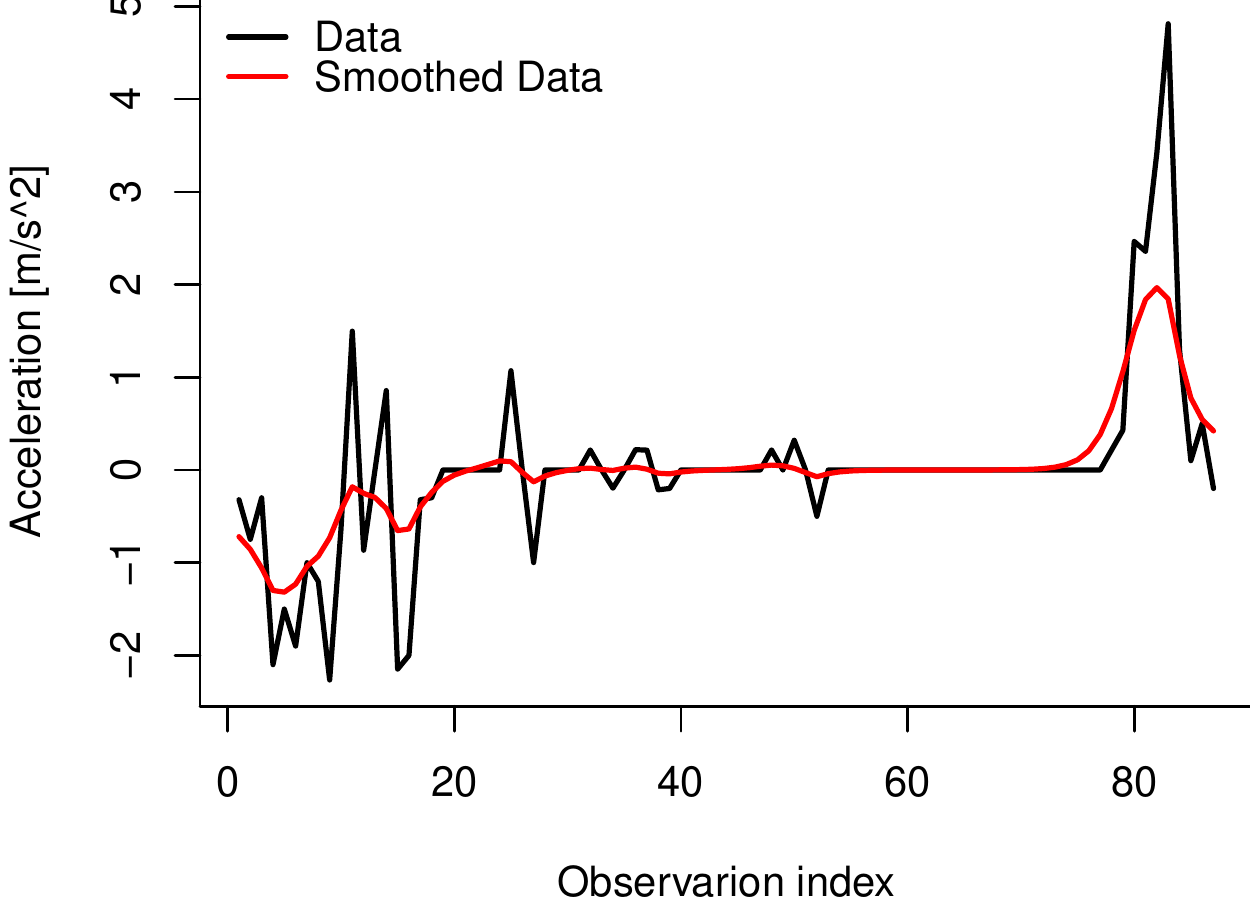} & \includegraphics[width=0.45\linewidth]{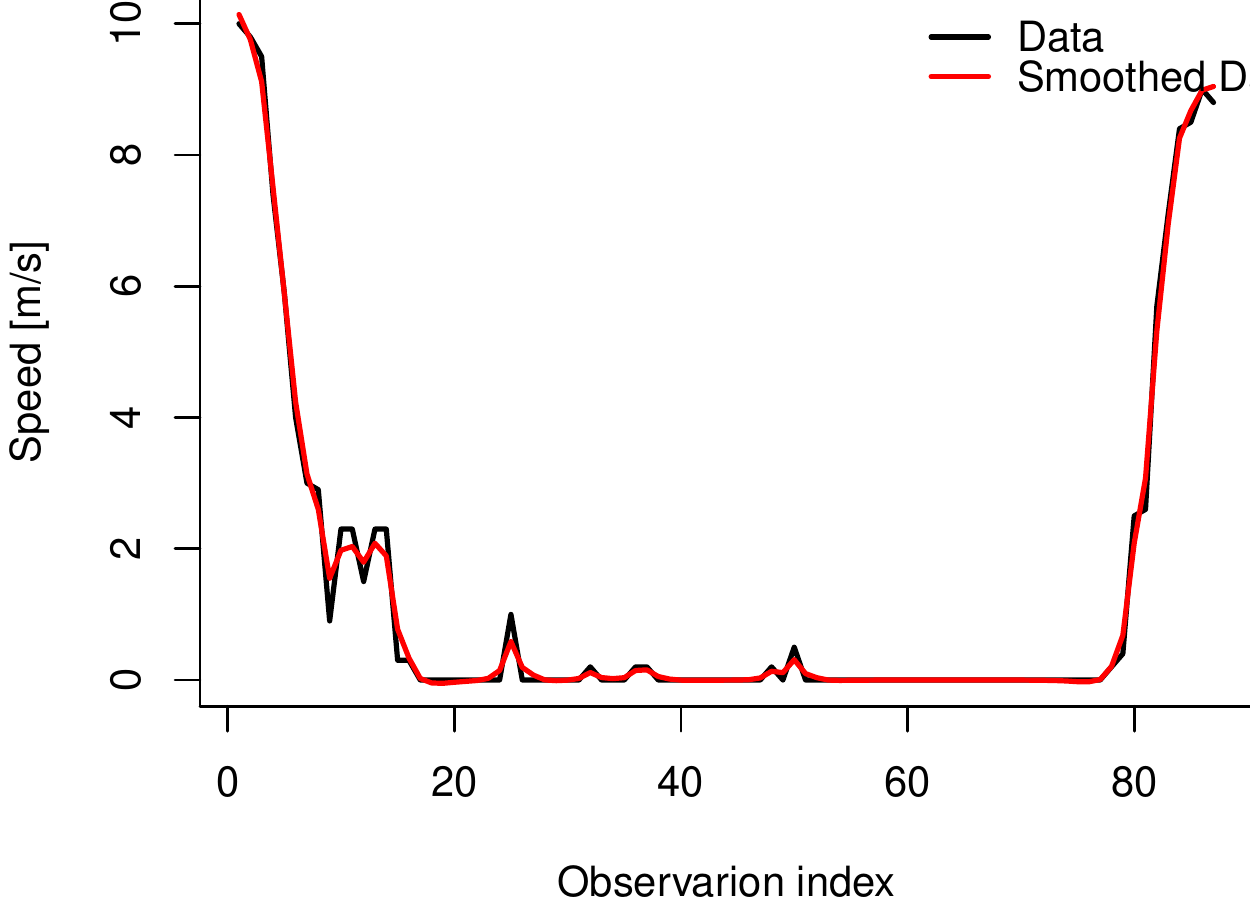}
	\end{tabular}
	\caption{Observations Smoothed with Kalman Filter}
	\label{fig:kalman}
\end{figure}

We formulate dynamics of the speed and acceleration observations as a state-space model
\begin{align}
y_t & = F\theta_t + v,~~v_t\sim N(0,V)\\
\theta_t & =  G\theta_{t-1} + w,~~ w\sim N(0,W)
\end{align}
Where $\theta_t = (s_t, a_t)$ vector with speed $s_t$ and acceleration $a_t$. Further, 
\[
F = \left(\begin{array}{cc}
1 & 0 \\ 
0 & 1
\end{array} \right),~~ 
G = \left(\begin{array}{cc}
1 & \Delta t \\ 
0 & 1
\end{array} \right)
\]
We used 
\[
V = \left(\begin{array}{cc}
%0.5 & 0 \\ 
0 & 1
\end{array} \right),~~ 
W = \left(\begin{array}{cc}
1 & 0 \\ 
0 & 0.2
\end{array} \right)
\]
Thus, we consider that change in speed from time step to time step as normally distributed with mean 0 and variance $0.5$ and change in acceleration follows a normal distribution with mean 0 and variance 1.  For each of the trajectories, we used $\mu_0 = (0,2)$ and $C_0 = 10^3\times\mathrm{diag}(2,2)$. Figure~\ref{fig:kalman} shows the result of applying Kalman smoothing (red line) to noisy speed and acceleration signals from phone GPS sensor (black line).

\subsubsection{Dynamic Time Warping}
Signals collected from CAN did not have associated locations. To add location attribute to each of the CAN signals, we joined CAN records with the phone records, which do have location attributes. Since the data was collected at different frequencies, a simple data base join operation that looks for equal time stamps would not work. We used dynamic time warping (DTW)~\cite{muller2007dynamic} to join the CAN and phone datasets. DTW finds an optimal alignment between two time series datasets. It ``warps'' one of the sequences to match the other one. Given two time-dependent sequences $a = (a_1,\ldots,a_N)$ and $y=(y_1,\ldots, y_n)$, DTW calculates the optimal warping path that minimizes the total distance between time series
\[
\rho_p(a,b) = \sum_{l=1}^{L}c(a_{p_a(l)},y_{p_b(l)}) \rightarrow \operatorname{minimize}_{p},
\]
here $p$ is the path function $p(l)  = (p_a(l),p_b(l))$ which for each index $l$ calculates corresponding indexes in $a$ and $b$ vectors. In our application the cost function $c(x,y) = |\operatorname{time}(x) - \operatorname{time}(y)|$ is the distance between the time stamps of each individual observations. 

\section{Analysis of Information Effect on Driving Behavior}
We applied data cleaning and filtering techniques described in the previous section to the datasets. Further, to remove bias, we only compared records from those roads where traffic signal information was available. The resulting cleaned phone speed data set contains 1,093,506 active observations and 1,311,666 inactive observations. An active observation was recorded when information about traffic signal timing was provided to the driver and inactive observations were recorded when no such information was provided.

First, we compare summary statistics for both active and inactive groups. The means for acceleration values are given in Table~\ref{tab:mean}.
\begin{table}[H]
	\centering
	\begin{tabular}{c|cc}
		Status & Mean positive acceleration &  Mean negative acceleration\\
		\hline
		Inactive & 0.7626 & -0.767\\
		Active & 0.746 & -0.741
	\end{tabular}
	\caption{Mean for positive and negative acceleration values}
	\label{tab:mean}
\end{table}

A one sided Welch two-sample $t$-test for the difference in the means $\mu_{\mathrm{inactive}} - \mu_{\mathrm{active}}$ confirms that the difference is significant. For positive acceleration the 95\% confidence interval for the difference is $(0.014,\infty)$ and for negative accelerations, interval is $(-\infty, -0.023)$ with $p$-value 10$^{-16}$ in both cases.

Further, we performed the $t$-test for the means of CAN signals.  Results of analysis of the CAN signals is shown in Table~\ref{tab:can_analysis}.

\begin{table}[H]
\begin{tabular}{l|llp{2cm}}
Signal & Inactive Mean & Active Mean & Conf. Interval\\\hline
Pedal D & 57.534 & 57.719 & (-0.225,-0.145)\\
RPM & 1285.8 & 1098.4 & (185.07,189.77)\\
Throttle & 50.769 & 48.862 & (1.8435,1.9692)\\
Acceleration Pedal & 55.944 & 62.959 & (-7.0984,-6.9321)\\
Throttle Relative & 20.458 & 17.207 & (3.185,3.3175)\\
Throttle Position & 79.808 & 90.074 & (-10.366,-10.166)\\
Fuel Rate & 9420.4 & 3222.3 & (5743.2,6653)\\
BMW RPM & 5183.5 & 5343.2 & (-165.73,-153.74)\\
BMW Fuel & 32708 & 32430 & (211.14,344.98)\\
\end{tabular}
\caption{Analysis of the means of CAN signals.}
\label{tab:can_analysis}
\end{table}
\vspace{-10pt}

We also performed $t$-test for individual road segments, however, even for the most traveled roads of the network, the sample sizes were not large enough to make conclusive statements. Table~\ref{tab:segment} shows the means and results of $t$-test for the top 10 most traveled roads in San Jose. 

\begin{table}[H]
\begin{tabular}{llll}
Road ID & Mean Inactive (\# obs) & Mean Active (\# obs) & $p$-value\\\hline
24166 & 0.826 (886) & 0.751 (2686) & 0.0013\\ 
77790 & 0.656 (1001) & 0.649 (1514) & 0.3904\\ 
22025 & 0.664 (813) & 0.668 (2733) & 0.5813\\ 
20566 & 0.626 (1392) & 0.738 (1595) & 1.0000\\ 
9028 & 0.722 (220) & 0.587 (440) & 0.0104\\ 
12686 & 0.708 (422) & 0.666 (1064) & 0.1287\\ 
80357 & 0.653 (3172) & 0.722 (257) & 0.9413\\ 
80356 & 0.494 (2054) & 0.677 (183) & 0.9999\\ 
9717 & 0.848 (655) & 0.922 (1162) & 0.9741\\ 
24552 & 0.510 (702) & 0.518 (2161) & 0.6511\\ 
\end{tabular}
\caption{Comparison of mean acceleration values for individual roads.}
\label{tab:segment}
\end{table}
\vspace{-10pt}

%> t.test(pgr1$accel[pgr1$accel>0 & pgr1$active==0],pgr1$accel[pgr1$accel>0 & pgr1$active==1],alternative = "greater")
%\begin{table}[H]
%	\centering
%	\begin{tabular}{c|cc}
%		Status & SD for positive acceleration &  SD negative acceleration\\
%		\hline
%		Inactive & 0.49 & 0.497\\
%		Active & 0.487 & 0.493
%	\end{tabular}
%	\caption{Standard Deviation for positive and negative acceleration values}
%	\label{tab:sd}
%\end{table}

The higher mean acceleration/deceleration values in the inactive group show that the information did have a positive effect. The difference in the means, though, can be result of many drivers to accelerating ``slightly faster'' without information or as a result of presence of sharp acceleration patterns in the data. To answer this question we need to analyze the sharp acceleration patterns. Statistically speaking we are interested in behavior of the extreme acceleration/deceleration observations, a.k.a. tail observations. 

\begin{figure}[h]
	\centering
	\begin{tabular}{cc}
		\includegraphics[width=0.45\linewidth]{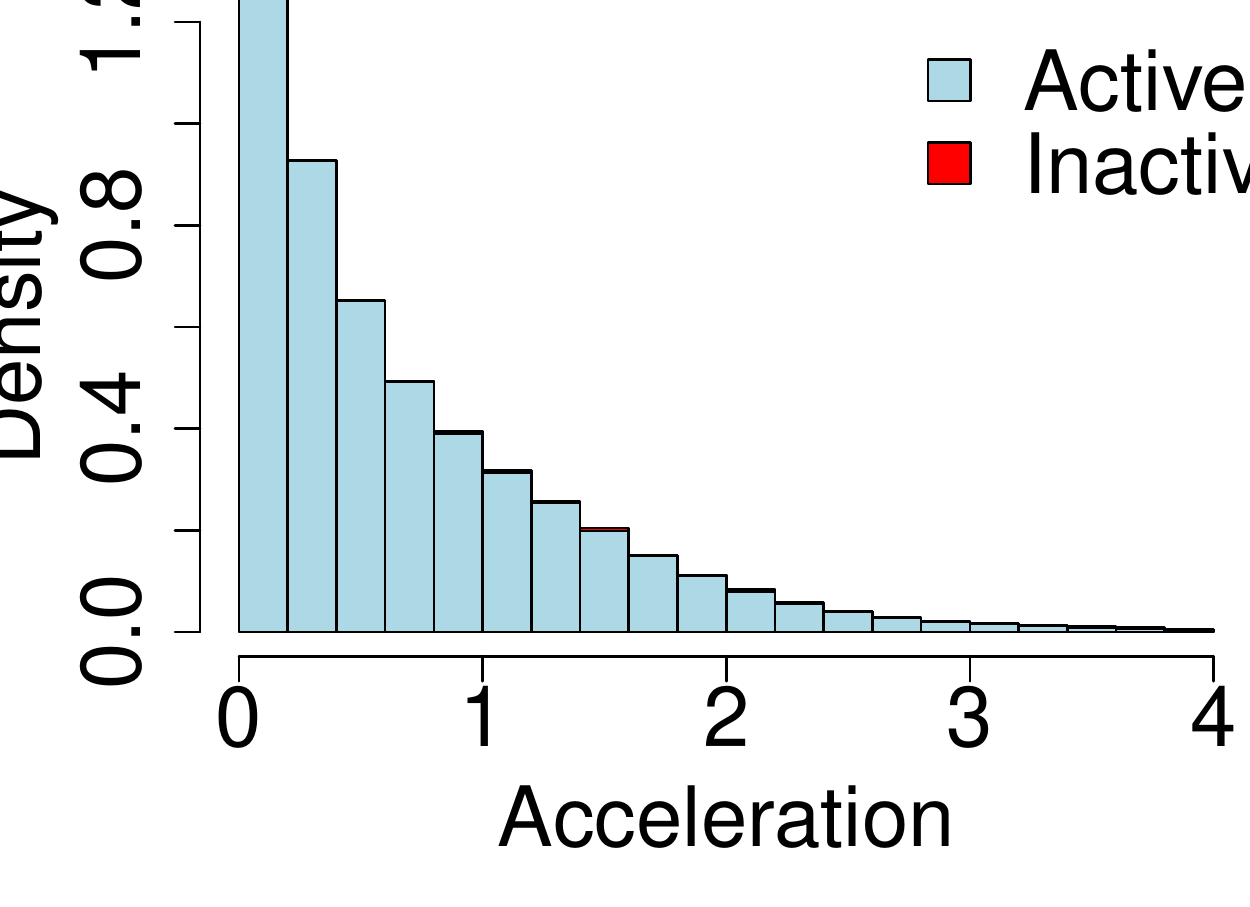} & \includegraphics[width=0.45\linewidth]{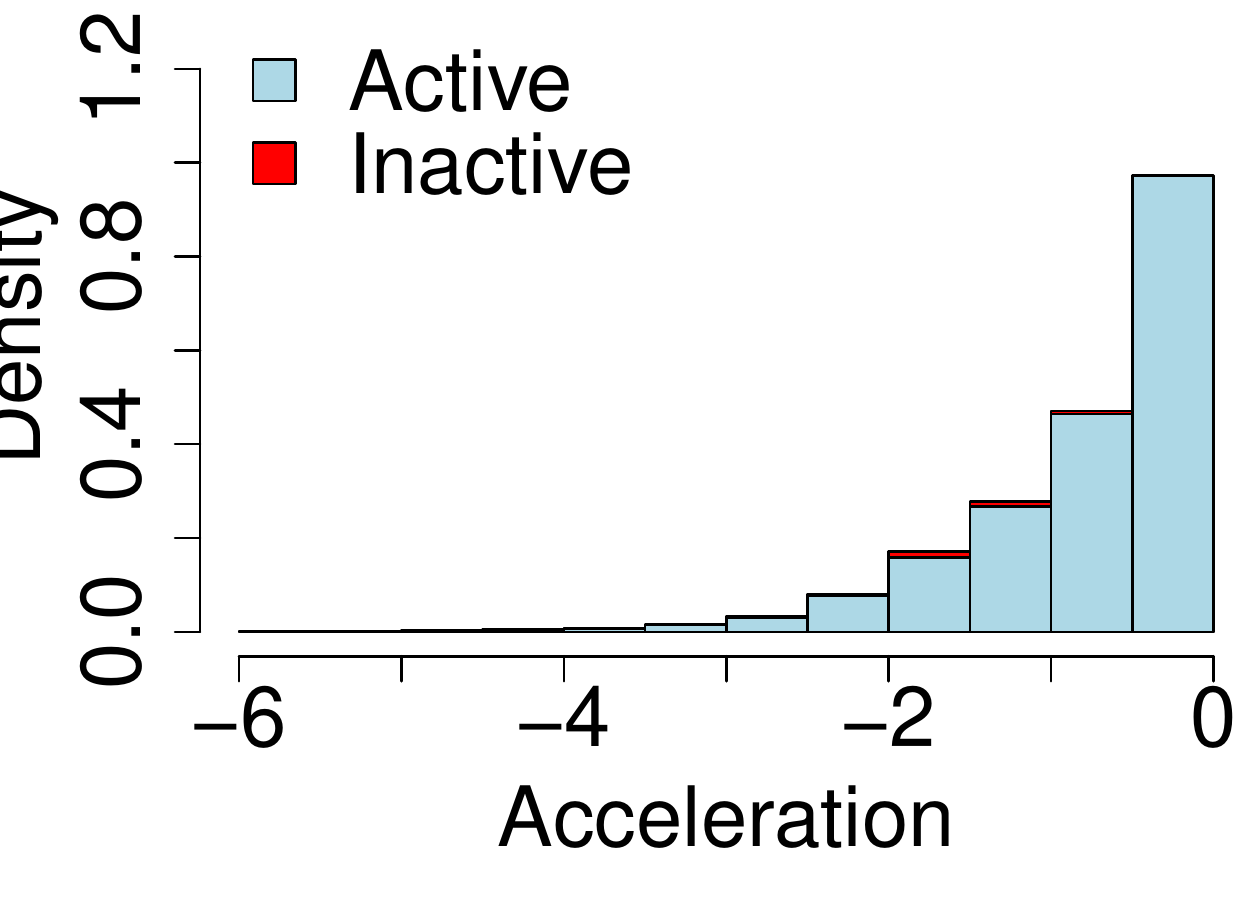}\\
		(a) Positive & (b) Negative
	\end{tabular}
	\caption{Histogram for (a) Positive and (b) Negative acceleration values}
	\vspace{-5pt}
	\label{fig:hist}
\end{figure}

Thus, instead of analyzing means, we need to analyze the entire distribution of the observations. Figure~\ref{fig:hist} shows the empirical distribution (histogram) for acceleration/deceleration for both active and inactive groups. 

\begin{figure}[h]
	\includegraphics[width=1\linewidth]{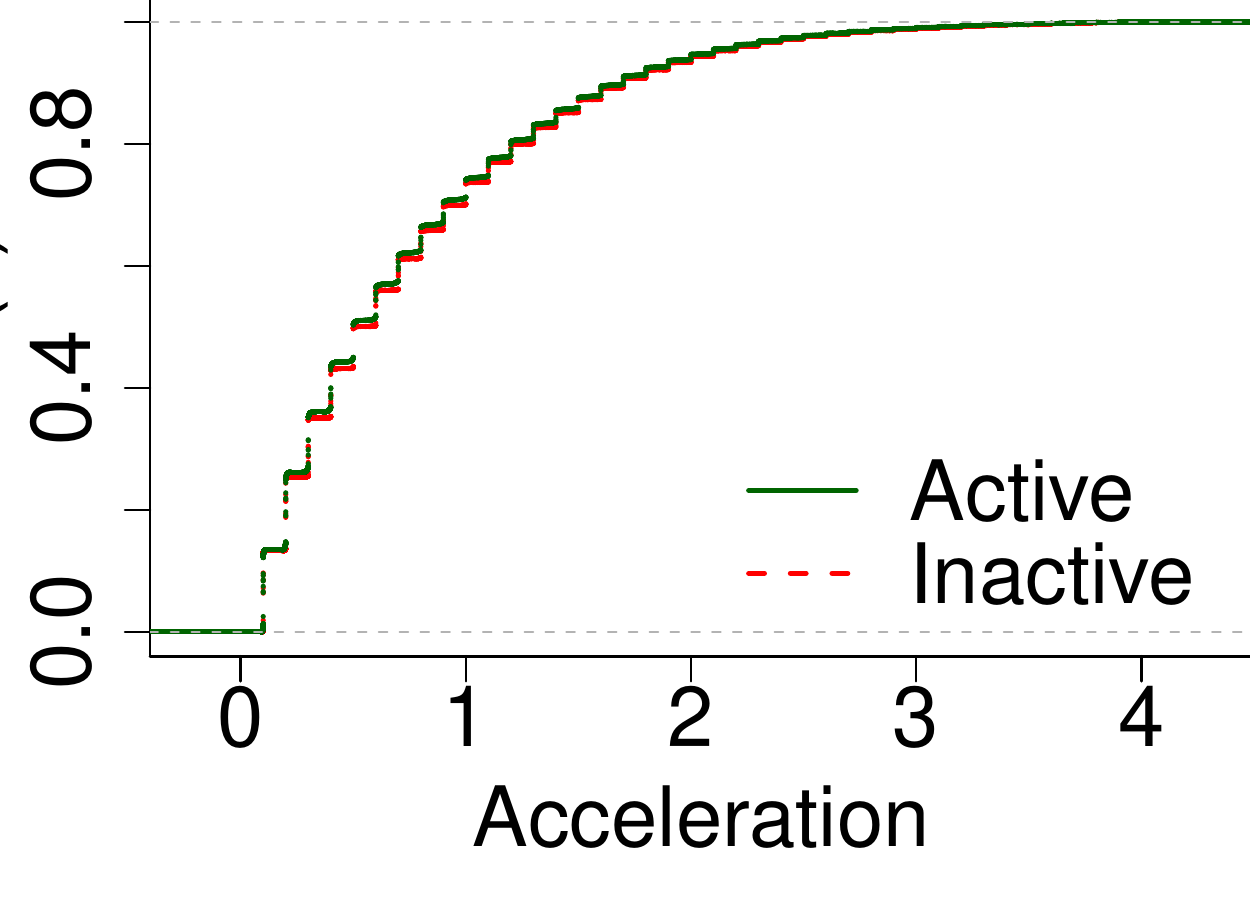}
	\caption{Empirical cumulative distribution for acceleration observations from active group (solid line) and inactive group (dashed line).}
	\vspace{-10pt}
	\label{fig:cdf-pos}
\end{figure}

Active values have more mass around modes of the distribution, while the distribution for inactive values has a heavier tail. The Kolmogorov-Smirnov test confirms the empirical observation that distributions are different. We perform a Kolmogorov-Smirnov test to compare empirical cumulative distributions.  Figure~\ref{fig:cdf-pos} shows the empirical cumulative distribution functions (CDF) for acceleration values for two groups of trips (active/inactive).

The Kolmogorov-Smirnov $D^-$-statistic equals 0.016 and the $p$-value is 10$^{-16}$. Thus, we accept the alternative hypothesis that the CDF of inactive values lies \textbf{below} that of active values for acceleration values.

\begin{figure}[h]
	\includegraphics[width=1\linewidth]{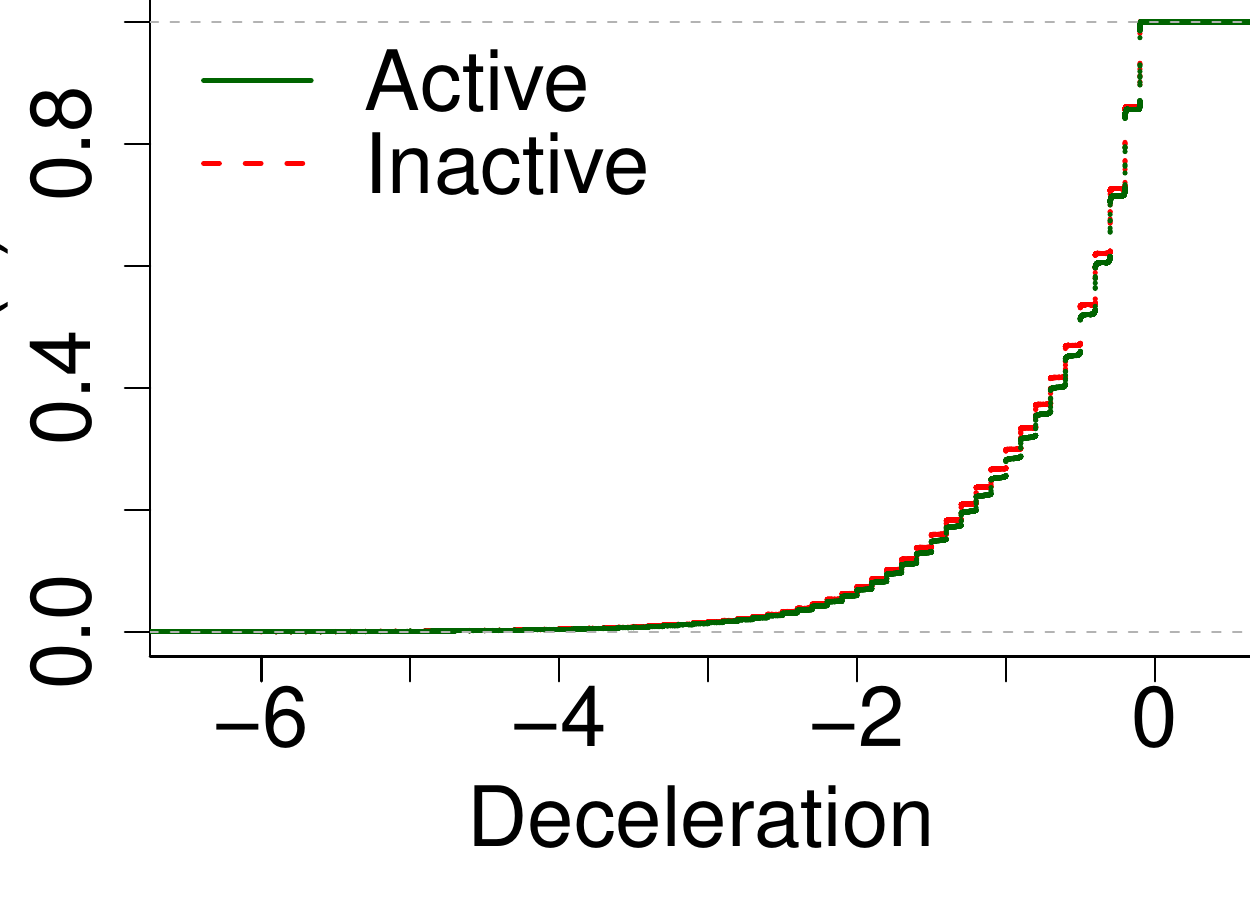}
	\caption{Empirical cumulative distribution for deceleration observations from active group (solid line) and inactive group (dashed line).}
	\vspace{-10pt}
	\label{fig:cdf-neg}
\end{figure}

Figure~\ref{fig:cdf-neg} shows the empirical cumulative distribution functions (CDF) for deceleration values for two groups of trips (active/inactive).

The Kolmogorov-Smirnov $D^+$-statistic equals 0.021 and the $p$-value is 10$^{-16}$. Thus, we accept the alternative hypothesis that the CDF of inactive values lies \textbf{above} that of active values for deceleration observations.

\subsection{Extreme Value Theory}
The Kolmogorov-Smirnov test confirms that the distributions of the active and inactive groups differ in their tails. To further quantify the difference of the tail observations, we use  \textit{Extreme value theory} (EVT) \cite{coles2001introduction}. EVT was developed to analyze extreme climate events and sharp market movements~\cite{smith2002measuring}. For example, Sigauke et. el. \cite{sigauke2013extreme} use EVT  for electricity demand modeling, and \cite{shenoy2014risk} provide a statistical model combined with EVT for electricity markets. 

We are interested in predicting the frequency at which a variable exceeds a certain threshold. For example, we consider acceleration greater than 3 m/s$^2$ to be aggressive driving and are interested in understanding the frequency of those events. Let $y$ denote our variable of interest, for example acceleration, and consider the exceedence over threshold events $\{y\mid y>u\}$. Then the probability of this event has a limiting generalized Pareto (GP) distribution, so that
\[
P(y \le a \mid y>u) \rightarrow H(a):=1-\left(1+\xi\dfrac{a-u}{\sigma}\right)^{-1/\xi}_+.
\]
Here  $(u,\sigma,\xi)$ are the location, scale and shape parameters,  $\sigma > 0$ and $z_+ = \max(z, 0)$. $H(a)$ is called Generalized Pareto (GP) distribution. The Exponential distribution is obtained by continuity as $\xi \rightarrow 0$. 

\begin{figure}[H]
\begin{tabular}{cc}
\includegraphics[width=0.45\linewidth]{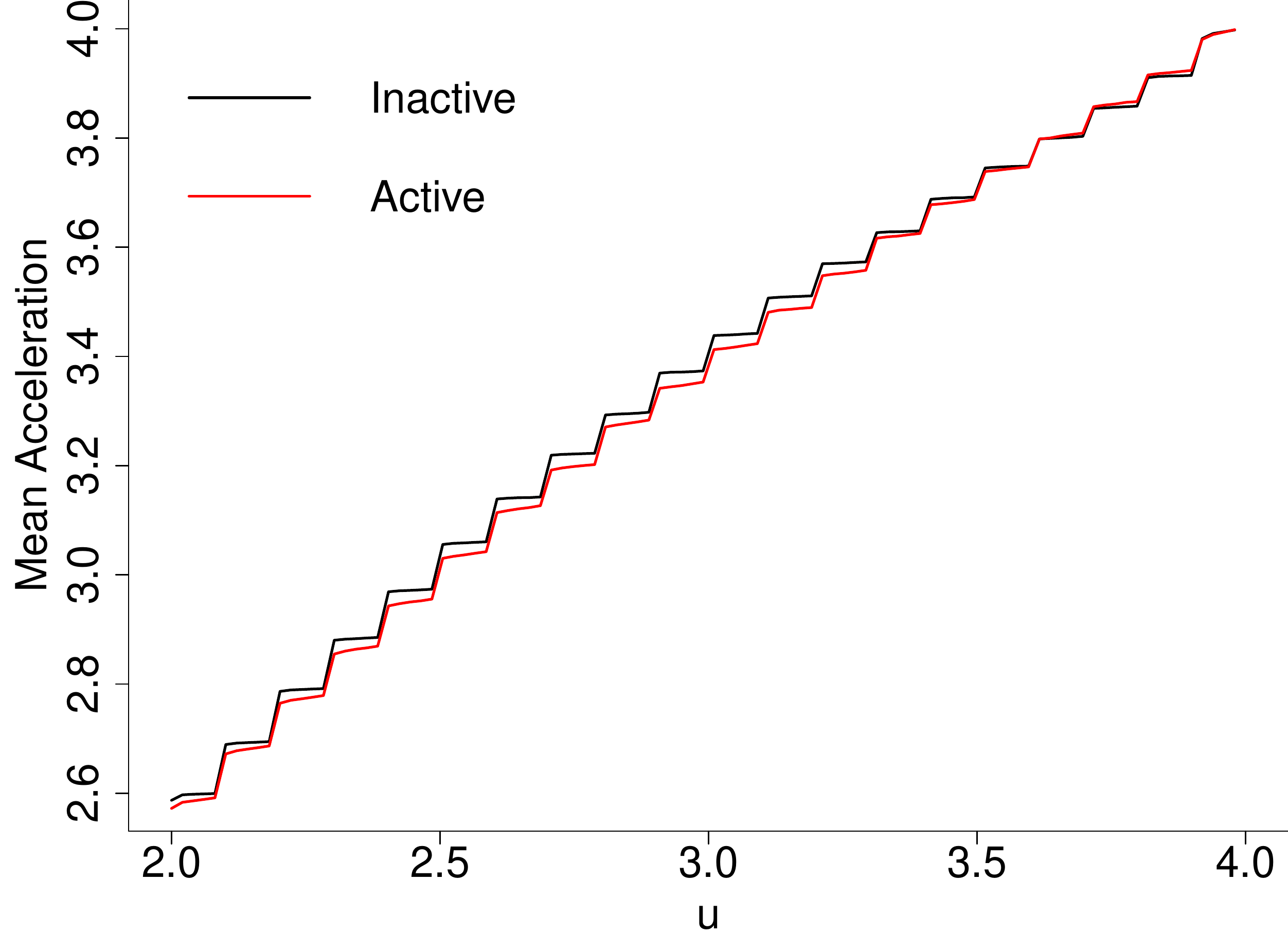} & \includegraphics[width=0.45\linewidth]{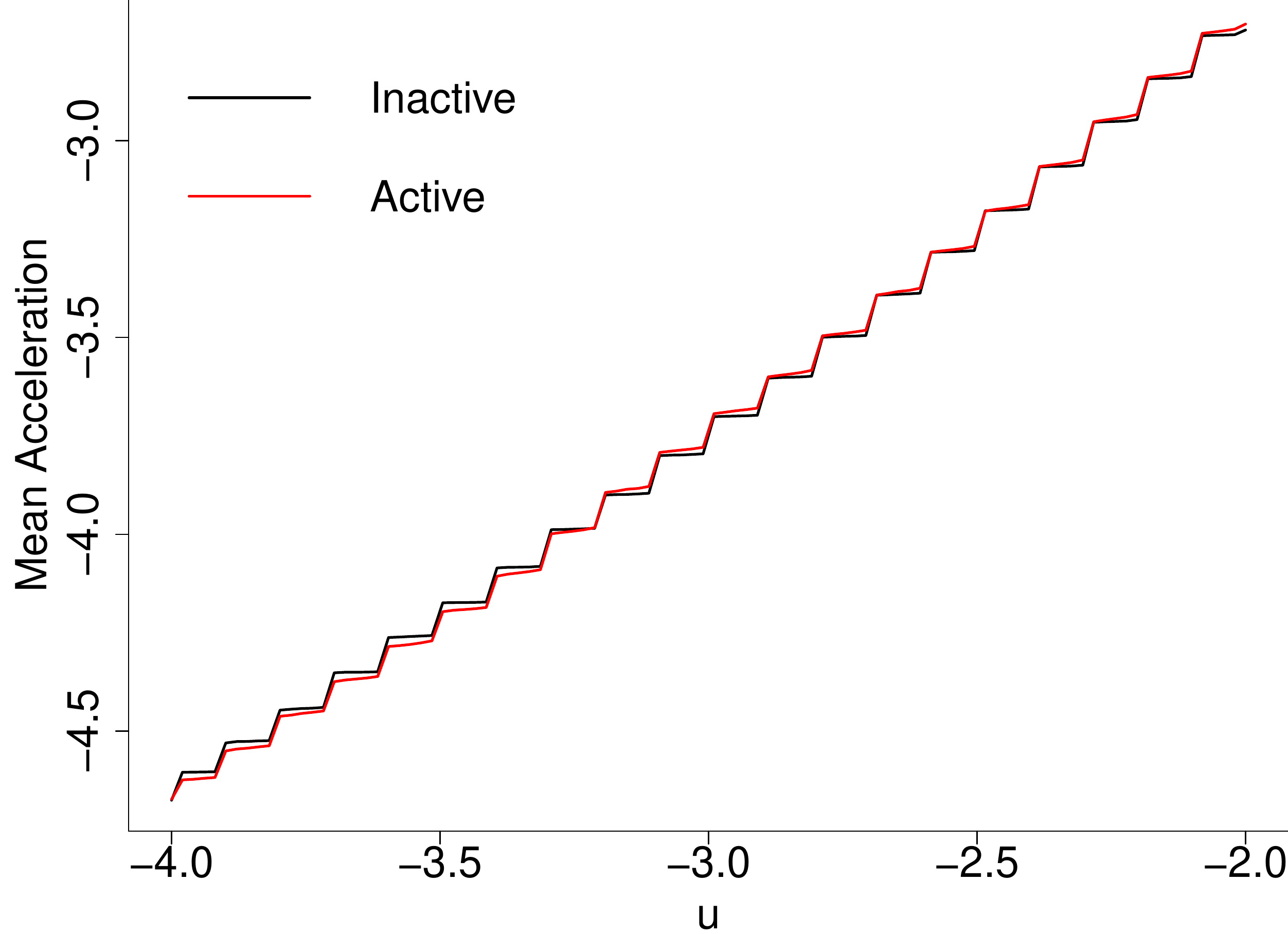}\\
(a) Acceleration & (b) Deceleration
\end{tabular}
\caption{Mean excess plot for active and inactive trips.}
\label{fig:mean_excess}
\end{figure}

To verify that this theoretical result holds for a given dataset, we can use the following property of the excedence function, empirically  if $y$ follows Generalized Pareto distribution, then for $\xi <1$ and $u>0$, we have
\[
E(y - u \mid y>u) = \dfrac{\sigma + \xi u}{1-\sigma}.
\]
An empirical plot of mean excess threshold should, therefore, be close to straight line with slope $\xi/(1-\xi)$. Figure~\ref{fig:mean_excess} shows that for our dataset this identity holds. Thus, we can use GP distribution to analyze the tails behavior.

\vspace{-10pt}

We fit the GP distribution and use it to calculate what threshold is expected to be exceeded every 24 seconds. Table~\ref{tab:gp_return} provides the result.
\begin{table}[H]
	\centering
\begin{tabular}{c|cc}
	&  Acceleration & Deceleration\\\hline
	Active & 2.18& -2.26\\
	Inactive &2.36 &-2.31 
\end{tabular}
\caption{Level exceeded every 24 seconds of the observation}
\label{tab:gp_return}
\end{table}
\vspace{-10pt}

The formal analysis of tails using GP distribution confirms our empirical observation that the inactive group has heavier tails. 

%Another way to model extreme values is through a point process, which jointly models the frequency of excedence and the time~\ref{leadbetter2012extremes}. In the absence of clustering (iid data) the number of observations above threshold $u$ during the time period $t_1,t_2$ has intensity that is follows Poisson distribution
%\[
%\Lambda(A) = (t_2-t_1)\left(1+\xi\dfrac{u-\mu}{\sigma}\right)^{-1/\xi}+,
%\]
%where $A = (t_1,t_2)\times (u,+\infty)$, $u$, $\sigma$ and $\xi$ are interpreted the same as in GP model. 

\section{Discussion}
The main contribution of this paper is the development and application of statistical techniques for analysis of driving data collected from smartphone and vehicle sensors. We analyzed two groups of observations: active and inactive. Active users received traffic information and inactive useres did not. We compared the means of the acceleration/deceleration observations and found that active drivers have smoother driving patterns. Our analysis demonstrates that there is a positive effect of providing traffic signal information timing to the drivers. Further, we analyzed extreme acceleration and deceleration patterns (tail observation). We showed that extreme acceleration and decelerations arise less frequently in the active group. The difference in the mean observations is not large. For example mean acceleration among active group is $0.76$26 m/s$^2$ and it is $0.746$ m/s$^2$ for the inactive group, which is a 2.2\% reduction. However, the difference among extreme accelerations/decelerations is more pronounced. We observed that every 24 seconds $2.18$ m/s$^2$ is expected to be exceeded by active group and $2.36$ m/s$^2$ for the inactive group. Thus, the reduction is 7.6\%.

There are many directions for future research. These include development of Bayesian analysis techniques which allow to ``pool'' data from multiple regions to derive metrics specific to individual road segments or intersections. Further, development of statistical models that use type of information provided as inputs would support understanding how drivers react to different types of messages. It will be useful to understand how changes in driving behavior impact vehicle emissions and safety metrics. For example, will smoother driving cycles lead to less accidents at the intersections and how will this impact fuel consumption and CO$_2$ emissions?

Finally, the data presented here are dependent on the level of accuracy of Connected Signals' predictive signal data and the particular presentation(s) of that data to drivers used for the study. More study on the effects of varying presentations and various accuracy requirements would clearly be beneficial.

\section{Acknowledgment}
The study is supported in part by the US Department of Energy under its ``Small Business Vouchers Pilot" program, and includes participation by Connected Signals, Inc., Argonne National Laboratories, BMW, and the cities of San Jose and Walnut Creek, California. 

\bibliography{ref}

\end{document}